\documentclass[twocolumn,showpacs,preprintnumbers,amsmath,amssymb,pra]{revtex4}
\usepackage{color}

\usepackage{graphicx}% Include figure files
\usepackage{dcolumn}% Align table columns on decimal point
\usepackage{bm}% bold math
\usepackage{epstopdf}
\usepackage{bbold}
\usepackage{amsthm}
\newtheorem{theorem}{Theorem}

\newtheorem{lemma}{Lemma}

\newcommand{\iden}{\mathbb{1}}
\newcommand{\diag}{\mathrm{diag}}
\newcommand{\zz}{\mathrm{Z}}
\newcommand{\tr}{\mathrm{Tr}}
\newcommand{\operation}{\mathcal{E}}

\newcommand{\TO}{\operation^\mathrm{TO}}
\newcommand{\STO}{\operation^\mathrm{STO}}
\newcommand{\ExTO}{\operation^\mathrm{ExTO}}

\newcommand{\TOcone}{\mathcal{C}^\mathrm{TO}}
\newcommand{\CTOcone}{\mathcal{C}^\mathrm{CTO}}
\newcommand{\STOcone}{\mathcal{C}^\mathrm{STO}}
\newcommand{\ExTOcone}{\mathcal{C}^\mathrm{ExTO}}

\begin{document}

\preprint{}

\title{Thermal Operations Involving a Single-mode Bosonic Bath}% Force line breaks with \\

\author{Xueyuan Hu}
\email{xyhu@sdu.edu.cn}
\affiliation{School of Information Science and Engineering, Shandong University, Qingdao 266237, China}%Lines break automatically or can be forced with \\

\date{\today}% It is always \today, today,

\begin{abstract}
We study the limitations on coherence evolutions under the constraints of thermodynamic laws, and focus on the optimal thermal operations (TO) reaching the bounds. For qubit case, we find a thermal operation involving only a single-mode reservoir (STO) which maintains the maximum coherence allowed by general TO. For higher dimensions, we derive general bounds on coherence merging under TO, and find STO to reach the bounds. By applying the bound to a two-qubit system, we prove that erasing correlations while preserving the marginal states is not free in the resource theory of thermodynamics. Due to the simple structure of STO and its strong ability in coherence processing, our results shed light on both theoretical and experimental studies in the field of thermodynamics for small quantum systems.
\end{abstract}

\maketitle

\section{Introduction}
Recently, the methods stemmed from quantum information theory have been applied to study the thermodynamics in the quantum and nano-scale regimes \cite{Nature2011NegEntropy,PhysRevLett.111.250404,PhysRevX.8.011019,PhysRevLett.120.040602,arXiv:1708.04302}. The main concern of this research field is to figure out the allowed state transformations under the constraints of thermodynamical laws. For this aim, one first needs to clarify the allowed operations. Such operations, known as thermal operations (TO) $\operation^\mathrm{TO}$ \cite{Horodecki2013thermalmajor}, are realized by the following general setting. A quantum system is brought into contact with a heat bath at an inverse temperature $\beta\equiv1/k_BT$, and then decoupled from the bath after some time. In some literatures, a catalytic system, which contacts with the system and then returns to its original state, is employed; such operations are called catalytic thermal operations (CTO) \cite{Brandao3275CTO}.% When the state of the system is diagonal in the energy eigenbasis, necessary and sufficient conditions are obtained for state transfer under TO \cite{Horodecki2013thermalmajor}, and well as under CTO \cite{Brandao3275CTO}.

By definition the heat bath is large, in the sense that both the energy and the size of degeneracies tend to infinity. Such structure of the heat bath causes difficulties in both theoretical analysis and experimental implementations. Hence, alternative thermal operations are proposed to simplify the analysis. For example, two important properties of TO, the time-translation symmetry and the Gibbs-preserving property, are notified. Each property has quite clean mathematical presentations. The operations which satisfy these two properties are called extended TO (ExTO) \cite{PhysRevLett.115.210403}. Besides, the experimental settings are proposed for the so-called elementary TO (ElTO) \cite{Lostaglio2018elementarythermal}, which are defined as TO involving only two energy levels. By definition, the comparisons between different sets of thermal operations are $\text{ElTO}\subsetneq\text{TO}\subseteq\text{ExTO}$ and $\text{TO}\subsetneq\text{CTO}$.

Because of the time-translation symmetry, quantum coherence between different energy levels can not be created by thermal operations, and is hence a resource. It attracts a lot of interest to study the dynamics of coherence under thermal operations \cite{PhysRevLett.115.210403,PhysRevX.5.021001,Science2003Scully,NC2015CoherenceLowTem,NC2015Lostaglio,NJP2016Korzekwa,PhysRevA.96.032109,arXiv:1302.2811,PhysRevLett.113.150402}. Despite that nice bounds have been derived for the output coherence after the action of ExTO \cite{PhysRevLett.115.210403,PhysRevX.5.021001}, it is difficult to verify whether those bounds are tight. This problem is only solved for very restricted cases, e.g., quantum operations involving only two energy levels \cite{PhysRevLett.115.210403}, and coherence shift-up and shift-down in a qutrit with nondegenerate energy levels and degenerate energy gaps \cite{PhysRevX.5.021001}.

In this paper, we propose a subset of thermal operations, the single-mode thermal operations (STO), which involve only a single-mode bosonic heat bath. In spite of the simple structure, STO perform well in processing coherence. We prove that for a single-qubit system, STO can induce any state transformations which are allowed by TO. Further, we derive the tight bounds on coherence merging under ExTO, between two energy gaps that does not share an overlap, and obtain the STO that achieve those bounds. The recently proposed optimal coherence shifting \cite{PhysRevX.5.021001} are special cases of our results. By applying the coherence merging bound to a two-qubit system, we find an example where it is impossible to transform a two-qubit state to the tensor product of the marginal states by thermal operations. It means that erasing the correlations between two quantum systems is not free in the resource theory of thermodynamics.

\section{Thermal operations and related concepts}
A completely positive and trace-preserving (CPTP) map is called a thermal operation $\TO$ if it can be written as \cite{Horodecki2013thermalmajor}
\begin{equation}
\label{eq:TO} \TO(\rho)=\tr_R(U[\rho\otimes\gamma_R]U^\dagger).
\end{equation}
Here $\gamma_R=e^{-\beta H_R}/\zz_R$ is a Gibbs state of the heat bath with Hamiltonian $H_R$ and partition function $\zz_R=\tr(e^{-\beta H_R})$, $\rho$ is a quantum state of a system with Hamiltonian $H$, and $U$ is a joint unitary commuting with the total Hamiltonian of the system and heat bath $[U,H+H_R]=0$.

A thermal operation must satisfy the following two properties \cite{PhysRevLett.115.210403,PhysRevX.5.021001}.\\
(P1) $\{\TO\}$ are time-translation symmetric
\begin{equation}
\TO(e^{-iHt}\rho e^{iHt})=e^{-iHt}\TO(\rho) e^{iHt}.
\end{equation}
(P2) $\{\TO\}$ preserve the Gibbs state
\begin{equation}
\TO(\gamma)=\gamma.
\end{equation}
An operation that satisfies (P1) is called a symmetric operation (SO) \cite{PhysRevX.5.021001,RevModPhys.79.555,PhysRevA.90.062110}. An operation satisfying both (P1) and (P2) is called an extended thermal operation (ExTO) \cite{PhysRevLett.115.210403}. The elementary thermal operations (ElTO) \cite{Lostaglio2018elementarythermal}, defined as TO involving only two energy levels, are proposed for experimental convenience. By definition, $\mathrm{ElTO}\subset\mathrm{TO}\subseteq\mathrm{ExTO}\subset\mathrm{SO}$. In order to study the comparisons between different sets of thermal operations, the operation cone is proposed \cite{Lostaglio2018elementarythermal}. For a given state $\rho$, the TO cone $\TOcone(\rho)$ is defined as the set of all states that can be prepared from $\rho$ under the action of TO:
\begin{equation}
\TOcone(\rho)=\{\rho'|\rho'=\TO(\rho)\}.
\end{equation}
The cones for other sets of thermal operations are defined similarly.

We first consider the population dynamics under different sets of thermal operations. For a system in a state $\rho$, let $\boldsymbol p$ be a vector of the occupation probabilities $p_k=\langle k|\rho|k\rangle$ of energy levels $E_k$, the population dynamics induced by a CPTP map $\operation$ can be represented as
\begin{equation}
\rho'=\operation(\rho)\Rightarrow \boldsymbol{p'}=G\boldsymbol p,
\end{equation}
Here $G$ is a matrix of transition probabilities $G_{k'k}=p_{k'|k}\equiv\langle k'|\operation(|k\rangle\langle k|)|k'\rangle$ from energies $E_k$ to $E_{k'}$, and $\boldsymbol{p'}$ is a vector of occupation probabilities $p'_k=\langle k|\rho'|k\rangle$ for the output state $\rho'$. From (P2), the population dynamics $G$ induced by ExTO is a stochastic matrix that preserves the Gibbs distribution. Such matrices, also referred to as Gibbs-stochastic matrices, can be realized by thermal operations \cite{Horodecki2013thermalmajor,Korzekwa16}. Hence when only population dynamics is concerned, ExTO is equivalent to TO. When the dimension of system $d\geq3$, some of the population dynamics induced by TO can not be realized by a sequence of ElTO \cite{Lostaglio2018elementarythermal}.

The dynamics of coherence between energy levels depends on both initial coherence and transition probabilities. For a quantum state $\rho$ expanded in its energy eigenbasis $\rho=\sum_{i,j}\rho_{ij}|i\rangle\langle j|$, a mode of coherence is defined as an operator $\rho^{\omega}$ composed of coherence terms between degenerate gaps:
\begin{equation}
\rho^{\omega}=\sum_{i,j:E_i-E_j=\hbar\omega}\rho_{ij}|i\rangle\langle j|.
\end{equation}
By a symmetric operation, each mode in the initial state is independently mapped to the corresponding mode of the final state \cite{PhysRevX.5.021001}. The output coherence term after the action of a symmetric operation is bounded as
\begin{equation}
\label{eq:cohdyn}|\rho'_{ij}|\leq{\sum_{c,d}}^\prime|\rho_{cd}|\sqrt{p_{i|c}p_{j|d}}.
\end{equation}
where the primed sum $\sum'$ means the summation only over indices $c,d$ satisfying $E_c-E_d=E_i-E_j$. As found in \cite{PhysRevLett.115.210403}, $\ExTOcone(\rho)=\TOcone(\rho)$ if $\rho$ is a qubit state. However, for higher dimension systems, there are situations where the state transformations under TO can not reach the above bound. Hence the authors of Ref. \cite{PhysRevLett.115.210403} believe that ExTO can outperform TO in coherence processing.

\section{Single-mode thermal operation}
A single-mode thermal operation is a thermal operation involving only a single-mode bosonic heat bath. Mathematically, it is written as in Eq. (\ref{eq:TO}), where the Hamiltonian of heat bath reads
\begin{equation}
\label{eq:hamhb}H_R=\sum_{n=0}^\infty n\hbar\omega|n\rangle_R\langle n|.
\end{equation}
The thermal state of the heat bath is then $\gamma_R^{\mathrm{STO}}=\sum_{n=0}^\infty\gamma_n|n\rangle\langle n|$ with $\gamma_n=q^n(1-q)$ and $q\equiv e^{-\beta\hbar\omega}$.

By the energy-preserving condition, a single-mode heat bath can only affect the quantum system whose energy gaps equal to $\hbar\omega$ (multiplied by a positive integer), so STO is a strict subset of TO. Nevertheless, we will show that when the size of the system is small, the STO cone occupies a large fraction of the TO cone. Especially, STO outperforms ElTO greatly in coherence processing. Compared with the general heat bath, a single-mode bosonic bath is friendly to experimental implementations.

Now we consider a $d$-dimension quantum system with Hamiltonian
\begin{equation}
\label{eq:hamsys}H=\sum_{k=0}^{d-1} k\hbar\omega|k\rangle_S\langle k|.
\end{equation}
The joint unitary $U$ is in the block-diagonal form $U=\oplus_{j=0}^\infty U^{(j)}$, where each block $U^{(j)}$ lives in a subspace with total energy $j\hbar\omega$. In the basis $\{|k\rangle_S\otimes|n\rangle_R\}$, the block $U^{(j)}$ is written as
\begin{equation}
\label{eq:jointU} U^{(j)}=\sum_{k,k'=0}^{d_j-1}U_{kk'}^{(j)}|k,j-k\rangle\langle k',j-k'|,
\end{equation}
where the subspace dimension $d_j=\max(d,j+1)$. Defining vectors $\vec A_{k'k}\equiv(A_{k'k}^{n_0},\cdots,A_{k'k}^n,\cdots)$ with $A_{k'k}^n=\sqrt{\gamma_n}U_{k'k}^{(k+n)}$ and $n_0=\max(0,(k'-k))$, we obtain the transition probabilities
\begin{equation}
\label{eq:popudyns}p_{k'|k}=|A_{k'k}|^2,
\end{equation}
as well as the dynamics of coherence terms
\begin{eqnarray}
\label{eq:cohdyns}\langle i|\STO(|c\rangle\langle d|)|j\rangle&=&\vec A^*_{i|c}\cdot\vec A_{j|d},
\end{eqnarray}
where $i-c=j-d$.
%The above inequality is equivalent to Eq. (\ref{eq:cohdyn}). Hence, if the bound of output coherence is reached, the nonzero vectors should satisfy $\vec A_{k,k+\Delta}\propto\vec A_{k',k'+\Delta}$.

Some constraints on the transition probabilities $p_{k'|k}$ can be derived directly from Eq. (\ref{eq:popudyns}). For example, $p_{0|0}\geq 1-q$, and for $k'>k$, $p_{k'|k}\leq q^{k'-k}$. These constraints forbids the exchange of occupations of different energy levels. When only the energy levels $k$ and $k'$ are involved and the bound $p_{k'|k}= q^{k'-k}$ is reached, the corresponding STO is just a $\beta$-swap \cite{Lostaglio2018elementarythermal} $\beta^{(k,k')}$ between levels $k$ and $k'$.

By the Cauchy-Schwarz inequality, Eq. (\ref{eq:cohdyns}) leads to Eq. (\ref{eq:cohdyn}), and the equality holds when the joint unitary $U$ exists such that $\vec A_{i|c}\propto\vec A_{j|d}$ for $i-c=j-d$. This provides a simple way to judge whether the bound as in Eq. (\ref{eq:cohdyn}) can be reached by STO, and if yes, to derive the explicit form of STO that reaches the bound.

\subsection{Qubit case}

If we only consider the population dynamics of a qubit system, various sets of thermal operations become equivalent. We clarify this observation in the following lemma.

\begin{lemma}\label{lemma:popudyn}
Consider a qubit system with Hamiltonian $H=E|1\rangle\langle1|$. If the initial state is diagonal $\rho_\diag=\diag(p_0,p_1)$, then we have
\begin{equation}
\TOcone(\rho_\diag)=\CTOcone(\rho_\diag)=\STOcone(\rho_\diag)=\ExTOcone(\rho_\diag).
\end{equation}
In the Bloch representation, $\TOcone(\rho_\diag)$ is the segment connecting the Bloch vectors of $\rho_\diag$ and $\beta^{0,1}(\rho_\diag)$. The transition probability that can be induced by TO satisfies $p_{0|0}\in[1-e^{-\beta E},1]$.
\end{lemma}

When coherence dynamics is considered, both ExTO and TO are proved to be able to achieve the bound as in Eq. \ref{eq:cohdyn} for qubits states \cite{PhysRevLett.115.210403}. Here we prove a stronger result that STO can also reach this bound. Precisely, for any qubit state $\rho$ and any given transition probabilities $p_{0|0}$ and $p_{1|1}$ (which preserve the thermal state), we find a STO $\STO_\mathrm{c}$ where the coherence term of the output state $\rho'=\STO_\mathrm{c}(\rho)$ reaches the bound $|\rho'_{01}|=\sqrt{p_{0|0}p_{1|1}}|\rho_{01}|$. The joint unitary of $\STO_\mathrm{c}$ is in the form of Eq. (\ref{eq:jointU}) with $U^{(0)}=|00\rangle\langle00|$, and for $j\geq2$, $d_j=2$ and
\begin{eqnarray}
U^{(j)}_{00}=U^{(j)}_{11}=\left(\frac{p_{1|1}}{p_{0|0}}\right)^{\frac{j}{2}},\nonumber\\
U^{(j)}_{01}=-U^{(j)}_{10}=\sqrt{1-\left(\frac{p_{1|1}}{p_{0|0}}\right)^{j}}.
\end{eqnarray}
By the Gibbs-preserving condition $p_{1|1}=1-e^{\beta\hbar\omega}(1-p_{0|0})\in[0,p_{0|0})$ and $p_{0|0}\geq1-e^{\beta\hbar\omega}>0$, so the joint unitary $U$ with the above parameters always exists. Together with Lemma 1, we obtain the following theorem.
\begin{theorem}\label{th:qubit}
For any qubit state $\rho$,
\begin{equation}
\STOcone(\rho)=\ExTOcone(\rho).
\end{equation}
\end{theorem}
It means that for a qubit system, the state transition induced by ExTO, can be realized by STO. For a high-dimension system, if an extended thermal operation involves only two energy levels, then its action on states can also be realized by STO.

\subsection{High-dimension case}
Now we consider a qutrit sytem whose Hamiltonian is Eq. (\ref{eq:hamsys}) with $d=3$. The ranges of output populations are derived from Eq. (\ref{eq:popudyns}). For a diagonal input state $\rho$ with $p_0>p_1/q>p_2/q^2$, we find that its STO cone is a strict subset of ElTO cone, which is in turn subset to TO (see Fig. \ref{fig:cone}). Nevertheless, STO outperform ElTO in processing coherence. For ElTO, the coherence term of output state is damped as $|\rho'_{kk'}|\leq\sqrt{p_{k|k}p_{k'|k'}}|\rho_{kk'}|$ \cite{Lostaglio2018elementarythermal}. From Theorem \ref{th:qubit}, this bound can be reached by STO. Moreover, STO can perform coherence processing tasks such as coherence shifting, while ElTO can not.

\begin{figure}
\includegraphics[width=0.9\columnwidth]{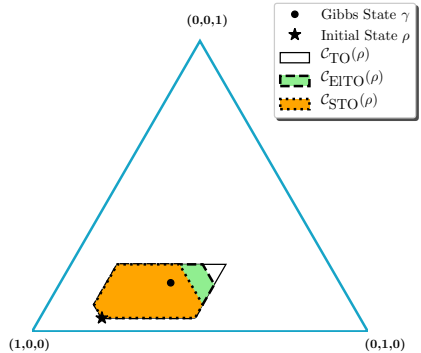}
\caption{\label{fig:cone}(color online). Comparison of the TO, ElTO, and STO cones for a qutrit state. The Hamiltonian of the qutrit is Eq. (\ref{eq:hamsys}) with $d=3$. The initial state is $\rho=\diag(0.8,0.16,0.04)$. Here the parameter $q\equiv e^{-\beta\hbar\omega}=0.5$.}
\end{figure}

An example was found where TO can not reach the coherence bound as in Eq. (\ref{eq:cohdyn}) (although it is still open whether the output coherence of TO can get arbitrary close to Eq. (\ref{eq:cohdyn})). In Appendix \ref{subsec:ExTOcohbound}, we prove that ExTO can always reach the bound. Hence TO is a strict subset of ExTO. We plot the comparison of different sets of thermal operations in Fig. 2.

\begin{figure}
\includegraphics[width=0.7\columnwidth]{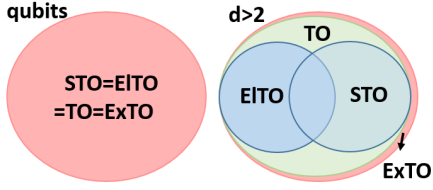}
\caption{\label{fig:cone}(color online). The comparison between ExTO, TO, STO, and ElTO.}
\end{figure}

\section{Coherence merging between degenerate gaps}
In a qubit system, the coherence term can not be increased by TO. From Eq. (\ref{eq:cohdyn}), in high-dimension systems, one coherence term $\rho_{ij}$ can be increased on the sacrifice of decreasing another coherence term $\rho_{cd}$ which is in the same mode as $\rho_{ij}$. Coherence merging is a task which merges many nonvanishing coherence terms in the same mode into a single coherence term. Precisely, let us consider a four-level system dominated by the following Hamiltonian:
\begin{equation}
H=\sum_{j=0}^3E_j|j\rangle\langle j|
\end{equation}
with $E_0=0$, $E_3=E_1+E_2$ and $E_1,E_2\geq0$. This Hamiltonian is nondegenerate but have degenerate gaps: $E_1-E_0=E_3-E_2$. It is easily identified that the mode $\omega_1\equiv E_1/\hbar$ is composed of two terms:
\begin{equation}
\rho^{\omega_1}=\rho_{10}|1\rangle\langle0|+\rho_{32}|3\rangle\langle2|.
\end{equation}
In the task of coherence merging, we want to merge them into a single coherence term $\rho'_{10}$ (or $\rho'_{32}$), i.e., to maximize $\rho'_{10}$ (or $\rho'_{32}$) for fixed $\rho_{10}$ and $\rho_{32}$. From Eq. (\ref{eq:cohdyn}),
\begin{equation}
\label{eq:cohmer}|\rho'_{10}|\leq\sqrt{p_{0|0}p_{1|1}}|\rho_{10}|+\sqrt{p_{0|2}p_{1|3}}|\rho_{32}|.
\end{equation}
For symmetric operations, the bound $|\rho_{10}|+|\rho_{32}|$ can be reached. The Kraus operators of such symmetric operation are $K_0=|0\rangle\langle0|+|1\rangle\langle1|$ and $K_1=|0\rangle\langle2|+|1\rangle\langle3|$. Nevertheless, this symmetric operation does not preserve the Gibbs state and is hence not a TO. Here we derive a tight bound for coherence merge under ExTO, which is stated as the following theorem. Detailed proof is in Appendix \ref{subsec:th2}.

\begin{theorem}\label{th:cohmer}
The bounds for coherence merging under ExTO are as follows. For coherence merging down,
\begin{eqnarray}
\label{eq:cohmerdown}|\rho'_{10}|\leq\max(|\rho_{10}|,(1-e^{-\beta E_2})|\rho_{10}|+|\rho_{32}|),
\end{eqnarray}
and for coherence merging up,
\begin{eqnarray}
\label{eq:cohmerup}|\rho'_{32}|\leq\max(|\rho_{10}|e^{-\beta E_2},|\rho_{32}|).
\end{eqnarray}
The above bounds can be achieved by a STO.
\end{theorem}

Depending on the ordering of $|\rho_{10}|e^{-\beta E_2}$ and $|\rho_{32}|$, the optimal TO that reach the above bounds is either identity or the quantum operation $\beta^{(0,2;1,3)}(\rho)$ which we call a simultaneous $\beta$-swap. The Kraus operators of $\beta^{(0,2;1,3)}$ are
\begin{eqnarray}
K_0^\beta&=&\sqrt{1-e^{-\beta E_2}}(|0\rangle\langle 0|+|1\rangle\langle1|),\nonumber\\
K_1^\beta&=&|0\rangle\langle2|+|3\rangle\langle1|,\\
K_2^\beta&=&\sqrt{e^{-\beta E_2}}(|2\rangle\langle 0|+|3\rangle\langle1|).\nonumber
\end{eqnarray}
The simultaneous $\beta$-swap $\beta^{(0,2;1,3)}$ can be realized by the STO with
\begin{eqnarray}
U&=&|00\rangle\langle00|+|10\rangle\langle10|\nonumber\\
&&+\sum_{n=1}^\infty(|0\rangle\langle2|+|1\rangle\langle3|)\otimes|n\rangle\langle n-1|\nonumber\\
&&+\sum_{n=1}^\infty(|2\rangle\langle0|+|3\rangle\langle1|)\otimes|n-1\rangle\langle n|,
\end{eqnarray}
and $H_R=\sum_{n=0}^\infty nE_2|n\rangle\langle n|$.

Let us analyse why the bound $|\rho_{10}|+|\rho_{32}|$ of coherence merging down can not be reached by a TO (the discussion for merging up is analogous). The effect of $K_1^\beta$ is to transfer the coherence term $\rho_{23}$ down, meanwhile, the populations on levels 2 and 3 are also transferred down. From the Gibbs-preserving condition, a transformation from levels 0 and 1 to levels 2 and 3 should be performed in order to balance the population. Such a transformation, which is realized by $K_2^\beta$, causes a decrease in the coherence term $\rho_{01}$. Therefore, we need to compare the amount of coherence merged down and the decrease caused in the coherence term $\rho_{01}$. If the former is larger, the TO for optimal coherence merging down is proved to be $\beta^{(0,2;1,3)}$; otherwise, the best way is to keep the state unchanged.

The four-level system discussed above can be considered as two qubits with Hamiltonian $H_A=E_2|1\rangle\langle1|$ and $H_B=E_1|1\rangle\langle1|$ respectively. In this regime, a direct consequence of Theorem \ref{th:cohmer} is that, it is not free in the resource theory of thermodynamics to decouple a composed system, i.e, to transform a composed state $\rho_{AB}$ to the tensor product of its marginal states $\rho_A\otimes\rho_B$ (where $\rho_A=\tr_B(\rho_{AB})$ and similar for $\rho_B$). To this end, let us consider the following two qubit state
\begin{equation}
\rho_{AB}=p|0\rangle^A\langle0|\otimes\rho_0^B+(1-p)|1\rangle^A\langle1|\otimes\rho_1^B,
\end{equation}
where the density matrices $\rho_{0,1}^B$ of qubit $B$ share the same diagonal elements $\langle0|\rho_0^B|0\rangle=\langle0|\rho_1^B|0\rangle$, but different off-diagonal elements $\langle0|\rho_0^B|1\rangle=a>0$, $\langle0|\rho_1^B|1\rangle=b>0$ and $a\neq b$. Direct calculations show that the coherence term of the tensor product state $\rho^A\otimes\rho^B$ is $\langle00|(\rho^A\otimes\rho^B)|01\rangle=p[pa+(1-p)b]$. Let $\rho'_{AB}=\ExTO(\rho_{AB})$ then from Theorem \ref{th:cohmer}, $\langle00|\rho'_{AB}|01\rangle\leq\max((1-q)pa+(1-p)b,pa)$. If $a<b<\frac{ap(p+q-1)}{(1-p)^2}$ (which implies $p>1/(1+q)$), then $\langle00|\rho'_{AB}|01\rangle<\langle00|(\rho^A\otimes\rho^B)|01\rangle$. In this case, $\rho^A\otimes\rho^B\notin\ExTOcone(\rho_{AB})$. It means that it is impossible to transform $\rho_{AB}$ to $\rho_A\otimes\rho_B$ by thermal operations.

Now we briefly discuss coherence merging between two coherence terms that have an overlap. Precisely, the system under consideration is a qutrit whose Hamiltonian is Eq. (\ref{eq:hamsys}) with $d=3$. The coherence terms $\rho_{10}$ and $\rho_{21}$ in the mode $\rho^\omega$ shares the same energy level 1. Let $|\rho_{10}|=a$ and $|\rho_{21}|=b$. Ref. \cite{PhysRevX.5.021001} gives the bounds for coherence merging under TOs as follows:
\begin{eqnarray}
|\rho'_{10}|&\leq&\sqrt{a^2+b^2},\ \ \text{for merging down},\\
|\rho'_{21}|&\leq&\sqrt{e^{-\beta\hbar\omega}a^2+b^2},\ \ \text{for merging up}.
\end{eqnarray}
As we have noticed, transferring the coherence term $\rho_{21}$ down must cause a decrease in $\rho_{10}$, and vise versa. Therefore, the above bounds are not tight. Here we prove tighter bounds for coherence merging under TOs:
\begin{eqnarray}
|\rho'_{10}|&\leq&\max\{\sqrt{(1-e^{-2\beta \hbar\omega})a^2+b^2},a\},\\
|\rho'_{21}|&\leq&\max\{ae^{-\beta \hbar\omega},b\}.
\end{eqnarray}

\section{conclusion}
We have studied the coherence evolutions under the constraints of thermodynamics, and focused on the optimal TO that can reach the bounds. Interestingly, for the case where only two energy levels are involved, \emph{and} for coherence merging between two coherence terms which do not share an energy level, we find that the bounds, which are derived for general TO, can be realized by thermal operations involving only a single-mode bosonic heat bath.

Besides, we find that it is not free in the resource theory of thermodynamics to transform a bipartite state to the tensor product of the marginal states by thermal operations. Hence, special attentions have to be paid when we employ a catalytic in a mixed state: not only the state of the catalytic must return to its original state, but also the correlation with other systems should be erased.

\begin{acknowledgments}
This work was supported by NSFC under Grant Nos. 11504205, 11774205, and Young Scholars Program of Shandong University.
\end{acknowledgments}

\newpage %Just because of unusual number of tables stacked at end
\bibliography{apssamp}% Produces the bibliography via BibTeX.

\begin{thebibliography}{20}
\expandafter\ifx\csname natexlab\endcsname\relax\def\natexlab#1{#1}\fi
\expandafter\ifx\csname bibnamefont\endcsname\relax
  \def\bibnamefont#1{#1}\fi
\expandafter\ifx\csname bibfnamefont\endcsname\relax
  \def\bibfnamefont#1{#1}\fi
\expandafter\ifx\csname citenamefont\endcsname\relax
  \def\citenamefont#1{#1}\fi
\expandafter\ifx\csname url\endcsname\relax
  \def\url#1{\texttt{#1}}\fi
\expandafter\ifx\csname urlprefix\endcsname\relax\def\urlprefix{URL }\fi
\providecommand{\bibinfo}[2]{#2}
\providecommand{\eprint}[2][]{\url{#2}}

\bibitem[{\citenamefont{Rio et~al.}(2011)\citenamefont{Rio, \AA{}berg, Renner,
  Dahlsten, and Vedral}}]{Nature2011NegEntropy}
\bibinfo{author}{\bibfnamefont{L.~d.} \bibnamefont{Rio}},
  \bibinfo{author}{\bibfnamefont{J.}~\bibnamefont{\AA{}berg}},
  \bibinfo{author}{\bibfnamefont{R.}~\bibnamefont{Renner}},
  \bibinfo{author}{\bibfnamefont{O.}~\bibnamefont{Dahlsten}}, \bibnamefont{and}
  \bibinfo{author}{\bibfnamefont{V.}~\bibnamefont{Vedral}},
  \bibinfo{journal}{Nature} \textbf{\bibinfo{volume}{474}}, \bibinfo{pages}{61}
  (\bibinfo{year}{2011}).

\bibitem[{\citenamefont{Brand\~ao et~al.}(2013)\citenamefont{Brand\~ao,
  Horodecki, Oppenheim, Renes, and Spekkens}}]{PhysRevLett.111.250404}
\bibinfo{author}{\bibfnamefont{F.~G. S.~L.} \bibnamefont{Brand\~ao}},
  \bibinfo{author}{\bibfnamefont{M.}~\bibnamefont{Horodecki}},
  \bibinfo{author}{\bibfnamefont{J.}~\bibnamefont{Oppenheim}},
  \bibinfo{author}{\bibfnamefont{J.~M.} \bibnamefont{Renes}}, \bibnamefont{and}
  \bibinfo{author}{\bibfnamefont{R.~W.} \bibnamefont{Spekkens}},
  \bibinfo{journal}{Phys. Rev. Lett.} \textbf{\bibinfo{volume}{111}},
  \bibinfo{pages}{250404} (\bibinfo{year}{2013}).

\bibitem[{\citenamefont{\AA{}berg}(2018)}]{PhysRevX.8.011019}
\bibinfo{author}{\bibfnamefont{J.}~\bibnamefont{\AA{}berg}},
  \bibinfo{journal}{Phys. Rev. X} \textbf{\bibinfo{volume}{8}},
  \bibinfo{pages}{011019} (\bibinfo{year}{2018}).

\bibitem[{\citenamefont{Lostaglio}(2018)}]{PhysRevLett.120.040602}
\bibinfo{author}{\bibfnamefont{M.}~\bibnamefont{Lostaglio}},
  \bibinfo{journal}{Phys. Rev. Lett.} \textbf{\bibinfo{volume}{120}},
  \bibinfo{pages}{040602} (\bibinfo{year}{2018}).

\bibitem[{\citenamefont{Gour et~al.}()\citenamefont{Gour, Jennings, Buscemi,
  Duan, and Marvian}}]{arXiv:1708.04302}
\bibinfo{author}{\bibfnamefont{G.}~\bibnamefont{Gour}},
  \bibinfo{author}{\bibfnamefont{D.}~\bibnamefont{Jennings}},
  \bibinfo{author}{\bibfnamefont{F.}~\bibnamefont{Buscemi}},
  \bibinfo{author}{\bibfnamefont{R.}~\bibnamefont{Duan}}, \bibnamefont{and}
  \bibinfo{author}{\bibfnamefont{I.}~\bibnamefont{Marvian}},
  \bibinfo{note}{arXiv:1708.04302}.

\bibitem[{\citenamefont{Horodecki and
  Oppenheim}(2013)}]{Horodecki2013thermalmajor}
\bibinfo{author}{\bibfnamefont{M.}~\bibnamefont{Horodecki}} \bibnamefont{and}
  \bibinfo{author}{\bibfnamefont{J.}~\bibnamefont{Oppenheim}},
  \bibinfo{journal}{Nature Communications} \textbf{\bibinfo{volume}{4}},
  \bibinfo{pages}{2059} (\bibinfo{year}{2013}).

\bibitem[{\citenamefont{Brand{\~a}o et~al.}(2015)\citenamefont{Brand{\~a}o,
  Horodecki, Ng, Oppenheim, and Wehner}}]{Brandao3275CTO}
\bibinfo{author}{\bibfnamefont{F.}~\bibnamefont{Brand{\~a}o}},
  \bibinfo{author}{\bibfnamefont{M.}~\bibnamefont{Horodecki}},
  \bibinfo{author}{\bibfnamefont{N.}~\bibnamefont{Ng}},
  \bibinfo{author}{\bibfnamefont{J.}~\bibnamefont{Oppenheim}},
  \bibnamefont{and} \bibinfo{author}{\bibfnamefont{S.}~\bibnamefont{Wehner}},
  \bibinfo{journal}{Proceedings of the National Academy of Sciences}
  \textbf{\bibinfo{volume}{112}}, \bibinfo{pages}{3275} (\bibinfo{year}{2015}).

\bibitem[{\citenamefont{\ifmmode \acute{C}\else
  \'{C}\fi{}wikli\ifmmode~\acute{n}\else \'{n}\fi{}ski
  et~al.}(2015)\citenamefont{\ifmmode \acute{C}\else
  \'{C}\fi{}wikli\ifmmode~\acute{n}\else \'{n}\fi{}ski,
  Studzi\ifmmode~\acute{n}\else \'{n}\fi{}ski, Horodecki, and
  Oppenheim}}]{PhysRevLett.115.210403}
\bibinfo{author}{\bibfnamefont{P.}~\bibnamefont{\ifmmode \acute{C}\else
  \'{C}\fi{}wikli\ifmmode~\acute{n}\else \'{n}\fi{}ski}},
  \bibinfo{author}{\bibfnamefont{M.}~\bibnamefont{Studzi\ifmmode~\acute{n}\else
  \'{n}\fi{}ski}}, \bibinfo{author}{\bibfnamefont{M.}~\bibnamefont{Horodecki}},
  \bibnamefont{and}
  \bibinfo{author}{\bibfnamefont{J.}~\bibnamefont{Oppenheim}},
  \bibinfo{journal}{Phys. Rev. Lett.} \textbf{\bibinfo{volume}{115}},
  \bibinfo{pages}{210403} (\bibinfo{year}{2015}).

\bibitem[{\citenamefont{Lostaglio et~al.}(2018)\citenamefont{Lostaglio,
  Alhambra, and Perry}}]{Lostaglio2018elementarythermal}
\bibinfo{author}{\bibfnamefont{M.}~\bibnamefont{Lostaglio}},
  \bibinfo{author}{\bibfnamefont{{\'{A}}.~M.} \bibnamefont{Alhambra}},
  \bibnamefont{and} \bibinfo{author}{\bibfnamefont{C.}~\bibnamefont{Perry}},
  \bibinfo{journal}{{Quantum}} \textbf{\bibinfo{volume}{2}},
  \bibinfo{pages}{52} (\bibinfo{year}{2018}), ISSN \bibinfo{issn}{2521-327X}.

\bibitem[{\citenamefont{Lostaglio
  et~al.}(2015{\natexlab{a}})\citenamefont{Lostaglio, Korzekwa, Jennings, and
  Rudolph}}]{PhysRevX.5.021001}
\bibinfo{author}{\bibfnamefont{M.}~\bibnamefont{Lostaglio}},
  \bibinfo{author}{\bibfnamefont{K.}~\bibnamefont{Korzekwa}},
  \bibinfo{author}{\bibfnamefont{D.}~\bibnamefont{Jennings}}, \bibnamefont{and}
  \bibinfo{author}{\bibfnamefont{T.}~\bibnamefont{Rudolph}},
  \bibinfo{journal}{Phys. Rev. X} \textbf{\bibinfo{volume}{5}},
  \bibinfo{pages}{021001} (\bibinfo{year}{2015}{\natexlab{a}}).

\bibitem[{\citenamefont{M.~O.~Scully and Walther}(2003)}]{Science2003Scully}
\bibinfo{author}{\bibfnamefont{G.~S.~A.} \bibnamefont{M.~O.~Scully},
  \bibfnamefont{M.~S.~Zubairy}} \bibnamefont{and}
  \bibinfo{author}{\bibfnamefont{H.}~\bibnamefont{Walther}},
  \bibinfo{journal}{Science} \textbf{\bibinfo{volume}{299}},
  \bibinfo{pages}{682} (\bibinfo{year}{2003}).

\bibitem[{\citenamefont{Narasimhachar and Gour}(2015)}]{NC2015CoherenceLowTem}
\bibinfo{author}{\bibfnamefont{V.}~\bibnamefont{Narasimhachar}}
  \bibnamefont{and} \bibinfo{author}{\bibfnamefont{G.}~\bibnamefont{Gour}},
  \bibinfo{journal}{Nature Communications} \textbf{\bibinfo{volume}{6}},
  \bibinfo{pages}{7689} (\bibinfo{year}{2015}).

\bibitem[{\citenamefont{Lostaglio
  et~al.}(2015{\natexlab{b}})\citenamefont{Lostaglio, Jennings, and
  Rudolph}}]{NC2015Lostaglio}
\bibinfo{author}{\bibfnamefont{M.}~\bibnamefont{Lostaglio}},
  \bibinfo{author}{\bibfnamefont{D.}~\bibnamefont{Jennings}}, \bibnamefont{and}
  \bibinfo{author}{\bibfnamefont{T.}~\bibnamefont{Rudolph}},
  \bibinfo{journal}{Nature Communications} \textbf{\bibinfo{volume}{6}},
  \bibinfo{pages}{6383} (\bibinfo{year}{2015}{\natexlab{b}}).

\bibitem[{\citenamefont{Korzekwa et~al.}(2016)\citenamefont{Korzekwa,
  Lostaglio, Oppenheim, and Jennings}}]{NJP2016Korzekwa}
\bibinfo{author}{\bibfnamefont{K.}~\bibnamefont{Korzekwa}},
  \bibinfo{author}{\bibfnamefont{M.}~\bibnamefont{Lostaglio}},
  \bibinfo{author}{\bibfnamefont{J.}~\bibnamefont{Oppenheim}},
  \bibnamefont{and} \bibinfo{author}{\bibfnamefont{D.}~\bibnamefont{Jennings}},
  \bibinfo{journal}{New Journal of Physics} \textbf{\bibinfo{volume}{18}},
  \bibinfo{pages}{023045} (\bibinfo{year}{2016}).

\bibitem[{\citenamefont{Lostaglio et~al.}(2017)\citenamefont{Lostaglio,
  Korzekwa, and Milne}}]{PhysRevA.96.032109}
\bibinfo{author}{\bibfnamefont{M.}~\bibnamefont{Lostaglio}},
  \bibinfo{author}{\bibfnamefont{K.}~\bibnamefont{Korzekwa}}, \bibnamefont{and}
  \bibinfo{author}{\bibfnamefont{A.}~\bibnamefont{Milne}},
  \bibinfo{journal}{Phys. Rev. A} \textbf{\bibinfo{volume}{96}},
  \bibinfo{pages}{032109} (\bibinfo{year}{2017}).

\bibitem[{\citenamefont{Skrzypczyk et~al.}()\citenamefont{Skrzypczyk, Short,
  and Popescu}}]{arXiv:1302.2811}
\bibinfo{author}{\bibfnamefont{P.}~\bibnamefont{Skrzypczyk}},
  \bibinfo{author}{\bibfnamefont{A.~J.} \bibnamefont{Short}}, \bibnamefont{and}
  \bibinfo{author}{\bibfnamefont{S.}~\bibnamefont{Popescu}},
  \bibinfo{note}{arXiv:1302.2811}.

\bibitem[{\citenamefont{\AA{}berg}(2014)}]{PhysRevLett.113.150402}
\bibinfo{author}{\bibfnamefont{J.}~\bibnamefont{\AA{}berg}},
  \bibinfo{journal}{Phys. Rev. Lett.} \textbf{\bibinfo{volume}{113}},
  \bibinfo{pages}{150402} (\bibinfo{year}{2014}).

\bibitem[{\citenamefont{Bartlett et~al.}(2007)\citenamefont{Bartlett, Rudolph,
  and Spekkens}}]{RevModPhys.79.555}
\bibinfo{author}{\bibfnamefont{S.~D.} \bibnamefont{Bartlett}},
  \bibinfo{author}{\bibfnamefont{T.}~\bibnamefont{Rudolph}}, \bibnamefont{and}
  \bibinfo{author}{\bibfnamefont{R.~W.} \bibnamefont{Spekkens}},
  \bibinfo{journal}{Rev. Mod. Phys.} \textbf{\bibinfo{volume}{79}},
  \bibinfo{pages}{555} (\bibinfo{year}{2007}).

\bibitem[{\citenamefont{Marvian and Spekkens}(2014)}]{PhysRevA.90.062110}
\bibinfo{author}{\bibfnamefont{I.}~\bibnamefont{Marvian}} \bibnamefont{and}
  \bibinfo{author}{\bibfnamefont{R.~W.} \bibnamefont{Spekkens}},
  \bibinfo{journal}{Phys. Rev. A} \textbf{\bibinfo{volume}{90}},
  \bibinfo{pages}{062110} (\bibinfo{year}{2014}).

\bibitem[{\citenamefont{Korzekwa}(2016)}]{Korzekwa16}
\bibinfo{author}{\bibfnamefont{K.}~\bibnamefont{Korzekwa}}, Ph.D. thesis,
  \bibinfo{school}{Imperial College London} (\bibinfo{year}{2016}).

\end{thebibliography}

\newpage
\section*{Appendices}
\subsection{Proof of Lemma 1.}
It is proven in Ref. \cite{Horodecki2013thermalmajor,Korzekwa16} that, any Gibbs-preserving population dynamics can be realized by TO. Therefore, $\ExTOcone(\rho_\diag)=\TOcone(\rho_\diag)$. From Lemma 1 of Ref. \cite{Lostaglio2018elementarythermal}, $\TOcone(\rho_\diag)=\STOcone(\rho_\diag)$.

For a qubit system, the population dynamics matrix induced by TO reads \cite{Lostaglio2018elementarythermal}
\begin{equation}
\label{eq:PDM}G^{(0,1)}=\left(\begin{array}{cc}
p_{0|0} & (1-p_{0|0})e^{\beta E}\\
1-p_{0|0} & 1-(1-p_{0|0})e^{\beta E}
\end{array}\right),
\end{equation}
where $p_{0|0}\in[1-e^{-\beta E},1]$ because all of the transition probabilities are nonnegative. For a state $\rho^\mathrm{TO}_\diag=\diag(p^\mathrm{TO}_0,1-p^\mathrm{TO}_0)\in\TOcone(\rho_\diag)$, we have
\begin{eqnarray}
p^\mathrm{TO}_0&=&p_{0|0}p_0+(1-p_{0|0})e^{\beta E}(1-p_0)\nonumber\\
&=&p_{0|0}[p_0-e^{\beta E}(1-p_0)]+e^{\beta E}(1-p_0).
\end{eqnarray}
Then $p_{0|0}\in[1-e^{-\beta E},1]$ is equivalent to that the value of $p^\mathrm{TO}_0$ is between $p_0$ and $p_0^{\mathrm c}\equiv1-p_0e^{-\beta E}$. Meanwhile, $\beta^{0,1}(\rho)=\diag(p_0^{\mathrm c},1-p_0^{\mathrm c})$. Hence, $\TOcone(\rho_\diag)$ the segment connecting the Bloch vectors of $\rho_\diag$ and $\beta^{0,1}(\rho_\diag)$.

Now we are left with the CTO cone. Because TO is a subset of CTO by definition, we only need to prove that if $\rho^\mathrm{CTO}_\diag=\diag(p^\mathrm{CTO}_0,1-p^\mathrm{CTO}_0)\in\CTOcone(\rho_\diag)$, then $p^\mathrm{CTO}_0$ is between $p_0$ and $p_0^{\mathrm c}$. From the free energy condition \cite{Brandao3275CTO}, $\rho_\diag$ can be transformed to a state arbitrarily close to $\rho^\mathrm{CTO}_\diag$ iff
\begin{equation}
D_\alpha(\rho||\gamma)\geq D_\alpha(\rho^\mathrm{CTO}||\gamma),\ \forall \alpha\in(-\infty,\infty),
\end{equation}
where $D_\alpha(\boldsymbol p||\boldsymbol q)=\frac{\mathrm{sgn}(\alpha)}{1-\alpha}\log{\sum_ip_i^\alpha q_i^{1-\alpha}}$ is the R{\'e}nyi divergences. For a diagonal qubit state $\sigma=\diag(r_0,r_1)$,
\begin{eqnarray}
D_\infty(\sigma||\gamma)&=&\log\zz+\log[\max\{q_0,q_1e^{\beta E}\}],\\
D_{-\infty}(\sigma||\gamma)&=&-\log\zz-\log[\min\{q_0,q_1e^{\beta E}\}].
\end{eqnarray}
If $\rho^\mathrm{CTO}_\diag\in\CTOcone(\rho_\diag)$, then $D_\infty(\rho_\diag||\gamma)\geq D_\infty(\rho^\mathrm{CTO}_\diag||\gamma)$ and $D_{-\infty}(\rho_\diag||\gamma)\geq D_{-\infty}(\rho^\mathrm{CTO}_\diag||\gamma)$, and in turn,
\begin{eqnarray}
\max\{p_0,p_1e^{\beta E}\}&\geq&\max\{p^\mathrm{CTO}_0,p^\mathrm{CTO}_1e^{\beta E}\},\\
\min\{p_0,p_1e^{\beta E}\}&\leq&\min\{p^\mathrm{CTO}_0,p^\mathrm{CTO}_1e^{\beta E}\}.
\end{eqnarray}
It means that $p^\mathrm{CTO}_0$ is between $p_0$ and $p_0^{\mathrm c}$. This completes the proof.

\subsection{Proof of Theorem 1.}
\begin{proof}
For a qubit system, joint unitary $U$ is in the block diagonal form $U=\oplus_{j=0}^\infty U^{(j)}$, where $U^{(0)}=|00\rangle\langle00|$ and for $j\geq1$, the diagonal blocks $U^{(j)}$ of are $2\times 2$ unitary operators, so we have $U^{(0)}_{00}=1$ and for $j\geq1$, $|U^{(j)}_{00}|=|U^{(j)}_{11}|$ and $|U^{(j)}_{01}|=|U^{(j)}_{10}|=\sqrt{1-|U^{(j)}_{00}|^2}$. Then from Eq. (\ref{eq:popudyns}), the transition probabilities are calculated as
\begin{eqnarray}
p_{0|0}&=&\frac{1}{\zz_R}\sum_{j=0}^\infty q^j|U^{(j)}_{00}|^2,\label{eq:p00}\\
p_{1|0}&=&\frac{1}{\zz_R}\sum_{j=1}^\infty q^j|U^{(j)}_{10}|^2=1-p_{0|0},\label{eq:p10}\\
p_{0|1}&=&\frac{1}{\zz_R}\sum_{j=1}^\infty q^{j-1}|U^{(j)}_{01}|^2=\frac{1}{q}p_{1|0}=\frac{1}{q}(1-p_{0|0}),\label{eq:p01}\\
p_{1|1}&=&\frac{1}{\zz_R}\sum_{j=1}^\infty q^{j-1}|U^{(j)}_{11}|^2\nonumber\\
&=&1-p_{0|1}=1-\frac{1}{q}(1-p_{0|0}).\label{eq:p11}
\end{eqnarray}
Hence, the transition probabilities $p_{1|0}$, $p_{0|1}$, and $p_{1|1}$, or equivalently the population dynamics, are determined by $p_{i|i}$. Because $U^{(0)}_{00}=1$ and $|U^{(j)}_{00}|\in[0,1],\ \forall j\geq1$, we have $p_{0|0}\in[1-q,1]$. From Lemma \ref{lemma:popudyn}, this is also the necessary and sufficient condition for a population dynamics achieve by ExTO.

Next, we consider the coherence dynamics enabled by a STO. For a qubit system, the coherence term can not increase under thermal operations. The damping factor $\eta$ of coherence is defined as $\big|\langle0|\STO(\rho)|1\rangle\big|=\eta|\rho_{01}|$. From Eq. (\ref{eq:cohdyns}), the damping factor of the coherence term reads
\begin{eqnarray}
\eta&=&\big|\langle0|\STO(|0\rangle\langle 1|)|1\rangle\big|=\vec A^*_{0|0}\cdot\vec A_{1|1}\nonumber\\
&\leq&|\vec A_{0|0}|\cdot|\vec A_{1|1}|=\sqrt{p_{0|0}p_{1|1}}
\end{eqnarray}
This is just the bound for processing coherence under ExTO (Eq.(\ref{eq:cohdyn})). Here the equality holds when $\vec A_{1|1}=\frac{|\vec A_{1|1}|}{|\vec A_{0|0}|}e^{i\varphi}\vec A_{0|0}$ for some $\varphi$. Thus,
\begin{eqnarray}
A_{11}^0=\sqrt{\frac{p_{1|1}}{p_{0|0}}}e^{i\varphi},\nonumber\\
A_{11}^j=(A_{11}^0)^je^{i\phi_{j-1}},\ A_{00}^j=(A_{11}^0)^je^{i\phi_{j}},\nonumber\\
A_{10}^j=\sqrt{1-|A_{11}^0|^{2j}}e^{i\psi_j},\nonumber\\
A_{01}^j=-\sqrt{1-|A_{11}^0|^{2j}}e^{i(\phi_j-\phi_{j-1}-\psi_j)}.
\end{eqnarray}
Because $0\leq p_{1|1}<p_{0|0}$, the above parameters in $U$ always exist. In a word, when only two energy levels are involved, both the population and the coherence dynamics induced by an extended TO can also be realized by a STO.
\end{proof}

\subsection{Proof of Theorem 2}\label{subsec:th2}
\begin{proof}
We prove the bound for merging down, and the results for merging up are analogous. By the Gibbs-preserving property, the transfering probabilities in Eq. (\ref{eq:cohmer}) should satisfy
\begin{eqnarray}
p_{0|0}+p_{0|2}e^{-\beta E_2}\leq 1,\\
p_{1|1}+p_{1|3}e^{-\beta E_2}\leq 1.
\end{eqnarray}
Labeling $p_{0|0}=\alpha_0^2\cos^2\theta_0,\ p_{0|2}e^{-\beta E_2}=\alpha_0^2\sin^2\theta_0,\ p_{1|1}=\alpha_1^2\cos^2\theta_1,\ p_{1|3}e^{-\beta E_2}=\alpha_1^2\sin^2\theta_1$ with $\alpha_0,\alpha_1\in[0,1]$ and $\alpha_j\cos\theta_j\in[\sqrt{\alpha_j^2-e^{-\beta E_2}},\alpha_j],j=0,1$, Eq. (\ref{eq:cohmer}) becomes
\begin{eqnarray}
|\rho'_{10}|&\leq&\alpha_0\alpha_1\cos\theta_0\cos\theta_1|\rho_{10}|\nonumber\\
&&+\alpha_0\alpha_1\sin\theta_0\sin\theta_1|\rho_{32}|e^{\beta E_2}\nonumber\\
&=&\alpha_0\alpha_1\cos(\theta_0-\theta_1)|\rho_{32}|e^{\beta E_2}\nonumber\\
&&+\alpha_0\alpha_1\cos\theta_0\cos\theta_1(|\rho_{10}|-|\rho_{32}|e^{\beta E_2})\nonumber\\
&\leq&\alpha_0\alpha_1|\rho_{32}|e^{\beta E_2}\nonumber\\
&&+\alpha_0\alpha_1\cos\theta_0\cos\theta_1(|\rho_{10}|-|\rho_{32}|e^{\beta E_2}).\label{eq:merdownbound}
\end{eqnarray}
If $|\rho_{10}|-|\rho_{32}|e^{\beta E_2}\geq0$, we obtain
\begin{eqnarray}
|\rho'_{10}|&\leq&\alpha_0\alpha_1|\rho_{32}|e^{\beta E_2}+\alpha_0\alpha_1(|\rho_{10}|-|\rho_{32}|e^{\beta E_2})\nonumber\\
&=&\alpha_0\alpha_1|\rho_{10}|\leq |\rho_{10}|.
\end{eqnarray}
It is easy to check that the equalities holds when $p_{j|j}=1,\forall j=0,1,2,3$. Hence, the identity operation or unitary operations diagonal in the energy eigenbasis can achieve this bound. Such operations can certainly be achieved by an STO.

For the case that $|\rho_{10}|-|\rho_{32}|e^{\beta E_2}<0$, Eq. (\ref{eq:merdownbound}) becomes
\begin{eqnarray}
|\rho'_{10}|&\leq&[\alpha_0\alpha_1-\sqrt{(\alpha_0^2-e^{-\beta E_2})(\alpha_1^2-e^{-\beta E_2})}]|\rho_{32}|e^{\beta E_2}\nonumber\\
&&+\sqrt{(\alpha_0^2-e^{-\beta E_2})(\alpha_1^2-e^{-\beta E_2})}|\rho_{10}|\nonumber\\
&\leq& (1-e^{-\beta E_2})|\rho_{10}|+|\rho_{32}|.\label{eq:merdownbound2}
\end{eqnarray}
The last inequality is from the fact that, the coefficient functions of both $|\rho_{32}|$ and $|\rho_{10}|$ maximize at $\alpha_0=\alpha_1=1$. We check that the population dynamics that reaches this bound reads
\begin{equation}
G=\left(\begin{array}{cccc}
1-e^{-\beta E_2} & 0 & 1 & 0\\
0 & 1-e^{-\beta E_2} & 0 & 1\\
e^{-\beta E_2} & 0 & 0 & 0\\
0 & e^{-\beta E_2} & 0 & 0
\end{array}\right).
\end{equation}
It is composed of two simultaneous $\beta-$swaps between levels 0,2 and between levels 1,3. Based on this consideration, we obtain the STO that reaches the bound in Eq. (\ref{eq:merdownbound2}): $\beta^{(0,2;1,3)}(\rho)=\tr_R[U(\rho\otimes\gamma_R)U^\dagger]$, where
\begin{eqnarray}
U&=&|00\rangle\langle00|+|10\rangle\langle10|\nonumber\\
&&+\sum_{n=1}^\infty(|0\rangle\langle2|+|1\rangle\langle3|)\otimes|n\rangle\langle n-1|\nonumber\\
&&+\sum_{n=1}^\infty(|2\rangle\langle0|+|3\rangle\langle1|)\otimes|n-1\rangle\langle n|,
\end{eqnarray}
and $\gamma_R=\exp(-\beta H_R)/\zz_R$ with $H_R=\sum_{n=0}^\infty nE_2|n\rangle\langle n|$. It can be checked that $U$ preserves the total energy of the system. Interestingly, the STO $\beta^{(0,2;1,3)}$ can also reach the coherence merging up bound for the case $|\rho_{10}|e^{-\beta E_2}\geq|\rho_{32}|$.
\end{proof}

\subsection{ExTO that reaches the bound of output coherence}\label{subsec:ExTOcohbound}
Here we derive the Kraus operators of the ExTO $\ExTO_{\mathrm c}$ which reaches the bound as in Eq. (\ref{eq:cohdyn}):
\begin{equation}
E^\Delta={\sum_k}^\Delta\sqrt{p_{k_\Delta|k}}|k_\Delta\rangle\langle k|,
\end{equation}
where $\Delta\in\{\Delta|\Delta=E_{k'}-E_{k},\ k,k'=0,\cdots,d-1\}$ is any possible gap, and the sum $\sum^\Delta_{k}$ means that when taking the summation over $k$, the index $k_\Delta$ should satisfy $E_{k_\Delta}=E_k+\Delta$.

The quantum operation $\sum_\Delta E^\Delta\cdot E^{\Delta\dagger}$ is a CPTP map, because
\begin{eqnarray}
\sum_\Delta E^{\Delta\dagger}E^\Delta&=&\sum_\Delta{\sum_{kk'}}^\Delta\sqrt{p_{k_\Delta|k}p_{k'_\Delta|k'}}|k'\rangle\langle k'_\Delta|k_\Delta\rangle\langle k|\nonumber\\
&=&\sum_\Delta{\sum_{k}}^\Delta p_{k_\Delta|k}|k\rangle\langle k|,\nonumber\\
&=&\sum_{k}\sum_{k_\Delta}p_{k_\Delta|k}|k\rangle\langle k|=\iden.
\end{eqnarray}

The action of $\ExTO_{\mathrm c}$ on any input state $\rho$ then reads
\begin{eqnarray}
\ExTO_{\mathrm c}(\rho)&=&\sum_\Delta{\sum_{kk'}}^\Delta\sqrt{p_{k_\Delta|k}p_{k'_\Delta|k'}}|k_\Delta\rangle\langle k|\rho|k'\rangle\langle k_\Delta'|\nonumber\\
&=&\sum_{k_\Delta k_\Delta'}{\sum_{kk'}}'\sqrt{p_{k_\Delta|k}p_{k_\Delta'|k'}}\rho_{kk'}|k_\Delta\rangle\langle k_\Delta'|,
\end{eqnarray}
where the primed sum $\sum'_{kk'}$ denotes the summation only over indices $k,k'$ such that $E_{k_\Delta}-E_k=E_{k_\Delta'}-E_{k'}$. It is easily checked that $p_{k_\Delta|k}$ are the transition probabilities (from energy level $E_k$ to $E_{k_\Delta}$), because
\begin{eqnarray}
\rho'_{k_\Delta k_\Delta}=\sum_{k}p_{k_\Delta|k}\rho_{kk}.
\end{eqnarray}

Clearly, each coherence term of output state is
\begin{equation}
\rho_{k_\Delta k_\Delta'}={\sum_{kk'}}'\sqrt{p_{k_\Delta|k}p_{k_\Delta'|k'}}\rho_{kk'},
\end{equation}
which is just the upper bound as in Eq. (\ref{eq:cohdyn}).

\end{document}